\documentclass[useAMS,usenatbib,letterpaper]{mn2e}
\usepackage{psfig} 
\newcommand{\hii}{{H{\sc\,ii}}}

\newcommand{\oiii}{[O~{\sc iii}]~5007\AA}

\newcommand{\kms}{{\rm km~s}$^{-1}$}

\begin{document}
\title[High speed outflows driven by the 30 Doradus starburst]
{High speed outflows driven by the 30 Doradus starburst}
\author[Redman, Al-Mostafa, Meaburn \& Bryce]
{M.P. Redman$^1$, Z.A. Al-Mostafa$^{2,3}$, J. Meaburn$^2$ and M. Bryce$^2$\\
$^1$Department~of~Physics~and~Astronomy, University~College London,
Gower Street,London WC1E 6BT, UK\\
$^2$Jodrell Bank Observatory, University of Manchester, 
Macclesfield SK11 9DL, UK\\
$^3$King Abdulaziz City for Science and Technology, 
Astronomy and Geophysics Research Institute, P.O.Box 6086 Riyadh 11442,
Saudi Arabia}
\date{\today}
\pubyear{2002} 
\volume{000}
\pagerange{\pageref{firstpage}--\pageref{lastpage}}
\maketitle 
\label{firstpage}

\begin{abstract}
Echelle spectroscopy has been carried out towards a sample region of
the halo of the giant \hii\ region 30 Doradus in the Large Magellanic
Cloud. This new kinematical data is the amongst the most sensitive yet
obtained for this nebula and reveals a wealth of faint, complex high
speed features. These are interpreted in terms of localised shells due
to individual stellar winds and supernova explosions, and collections
of discrete knots of emission that still retain the velocity pattern
of the giant shells from which they fragmented. The high speed
velocity features may trace the base of the superwind that emanates
from the 30 Doradus starburst, distributed around the super star
cluster R136.
\end{abstract}


\section{Introduction}
The 30 Doradus nebula in the Large Magellanic Cloud (LMC) is the
closest example of a giant extragalactic \hii\ region and the largest
in the local group of galaxies. It is regarded as undergoing intense
enough star formation to be referred to as a `mini-starburst' by
\citet{leitherer98} and as such is an important nearby laboratory of
both massive star formation and starburst phenomena. The highly
dynamic nebulosity (e.g.~\citealt{meaburn81,meaburn87}) is powered by
a super star cluster of $\sim 100$ massive stars. Remarkable HST
imagery of the environment of the central cluster of massive stars has
recently been presented by \citet{walborn.et.al02}. The combined
winds, UV radiation and supernova explosions from so many massive
stars at a similar evolutionary epoch enables the generation of the
nested giant (20 -- 300 pc diameter) shells that comprise the giant
\hii\ region (\citealt{meaburn80,meaburn90,leitherer98}). On the largest 
scales, surrounding 30 Dor are supergiant (600 -- 1400 pc diameter)
interstellar shells such as LMC3.

The term 'shell' will be used in this paper rather than the commonly
used term `bubble' since it is preferable to use a term that is
dynamically neutral and constrained to no specific geometry
(e.g. spherical). The term `bubble', often erroneously presupposes a
roughly spherical, pressure--driven, energy-conserving shell. This is
certainly not the case for the supergiant shells which are unlikely to
be either spherical or energy-conserving. The division between 'giant'
and 'supergiant' when applied to the LMC shells will be for the
diameter ranges above and recently confirmed by the H~{\sc i}
observations of \citet{kim.et.al99}. Different, though related,
mechanisms must be involved in the formation of LMC shells in these
distinctly separate diameter ranges. The most important difference is
that supergiant shells have diameters in excess of the neutral gas
scale-height of the LMC.

The overlapping giant shells comprising the halo of 30 Doradus have
been shown to be expanding at around 50 \kms\
(e.g. \citealt{meaburn84,chu&kennicutt94}) whereas a multitude of
$15~{\rm pc}$ diameter regions exhibit outflows of $\ga 200~{\rm
km~s^{-1}}$. The latter were interpreted as young supernova remnants
in the perimeters of giant shells \citep{meaburn88}. Fig.\@~1 is a
cartoon that illustrates the hierarchy of scale sizes present in a
giant \hii\ region like 30 Doradus.

The brightest, dominant velocity components of 30 Doradus are complex
but seem to be comprised of three distinct velocity regimes
corresponding to H~{\sc i} sheets along the sightline. These are at
$250~{\rm km~s^{-1}}$, $270~{\rm km~s^{-1}}$ and $300~{\rm km~s^{-1}}$
\citep{mcgee.et.al78,chu&kennicutt94,kim.et.al99}. In this paper, the systemic
velocity, $V_{\rm sys}$ is taken to be the average heliocentric
velocity ($V_{\rm HEL}$) of these components, $270~{\rm km~s^{-1}}$,
in agreement with previous observations (see for example
\citealt{peck.et.al97short,meaburn91,garay.et.al93,clayton87}). 

In this work, the aim is to investigate the faint highest speed
phenomena in the halo of 30 Doradus in order to complete the
kinematical characterisation of this important giant \hii\ region.
New echelle observations of the line profiles of the highest speed
phenomena in the halo of 30 Doradus have been made with unprecedented
sensitivity. These are described in Section 2. The region of study was
in the vicinity of the Wolf-Rayet star R130 and was selected as a
representative portion of the halo of 30 Doradus. This area is located
in `shell 3', of one of the several giant shells that comprise the 30
Doradus region (nomenclature from \citealt{meaburn80,chu&kennicutt94},
see also \citealt{wang&helfand91}) and was one of the regions
investigated by \citet{chu&kennicutt94}. Section 3 is a discussion of
the high--speed motions and morphologies in terms of the structures
and dynamics of the giant shells that comprise 30
Doradus. How the high speed features relate to the outflow of hot gas
from the 30 Doradus nebula is also discussed. Conclusions are drawn in
Section 4.

\begin{figure}
\psfig{file=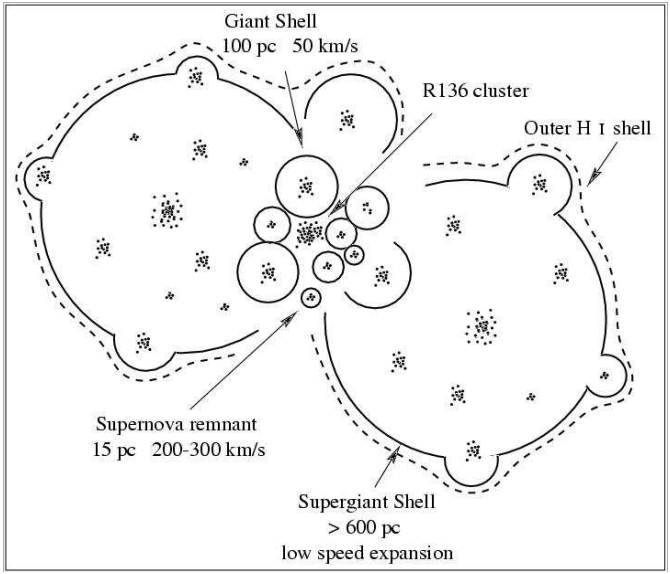,width=250pt,bbllx=20pt,bblly=20pt,bburx=575pt,bbury=500pt}
\caption{Cartoon of the 30 Doradus nebula to illustrate the hierarchy 
of shell scale sizes. The ambient density increases towards R136.}
\label{cartoon}
\end{figure}
  
\section{Observations and Results}
\begin{figure}
\psfig{file=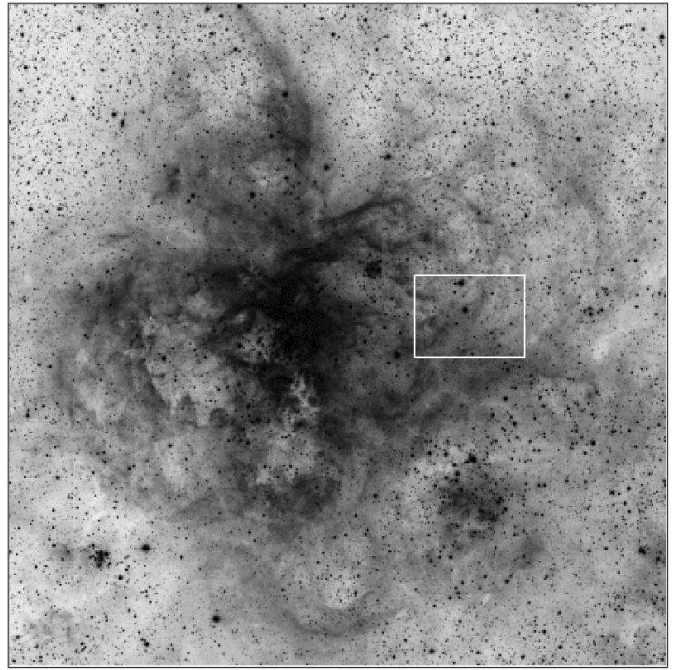,width=250pt,bbllx=20pt,bblly=20pt,bburx=575pt,bbury=584pt}
\caption{30 Doradus nebula. This figure is a cropped reproduction of ESO PR 
Photos 14a/02. The white box marks the region from which line profiles
were obtained. The size of the region displayed is approximately 200
pc across}
\label{eso30dor}
\end{figure}
\begin{figure}
\psfig{file=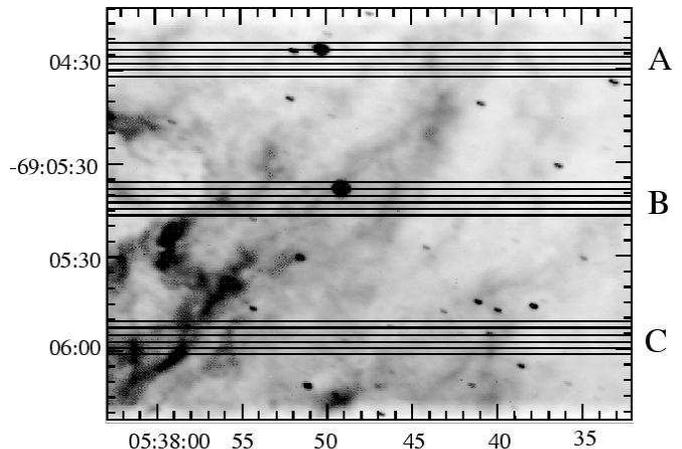,width=250pt,bbllx=20pt,bblly=20pt,bburx=575pt,bbury=395pt}
\caption{MES slit positions marked against the background nebulosity. This
region corresponds to the white box in the previous figure }
\label{slits}
\end{figure}
A reproduction of the ESO PR Photo 14a/02 is shown in Fig. 2. This is
an optical image of the 30 Doradus nebula obtained with the Wide Field
Imager camera on the 2.2-m MPG/ESO telescope at the La Silla
Observatory. A white box marks the region which is investigated
spectroscopically in the present work.

Spatially resolved, long-slit echelle spectra of \oiii\ emission lines
were obtained using the Manchester Echelle Spectrometer (MES)
\citep{meaburn.et.al84} at the 3.9M Anglo-Australian Telescope (AAT)
on the night of 1994 October 17. The data were obtained using three
parallel east-west slits simultaneously, separated by 45\arcsec\ which
were then stepped southwards in 2\arcsec\ increments. The slit length
was 165\arcsec\ and the slit width was $300~{\rm \mu m} (\equiv
2\arcsec $ and $20~{\rm km~s^{-1}}$). The broad slit compromises
spatial and spectral resolution but permits detection of the faintest,
highest speed components of the line profiles. The exposure time for
each multi-slit position was 1800s. The locations of all the slit
positions are shown in Fig. 3 and the boundary of this figure
corresponds to the white box in Fig. 2. In the discussion that
follows, a slit is identified by which block of six it appears
in. From north to south the blocks are labelled A, B, and C while
within a block, also from north to south, the slits are numbered from
1 to 6. The data were processed in the standard manner.

\begin{figure*}
\psfig{file=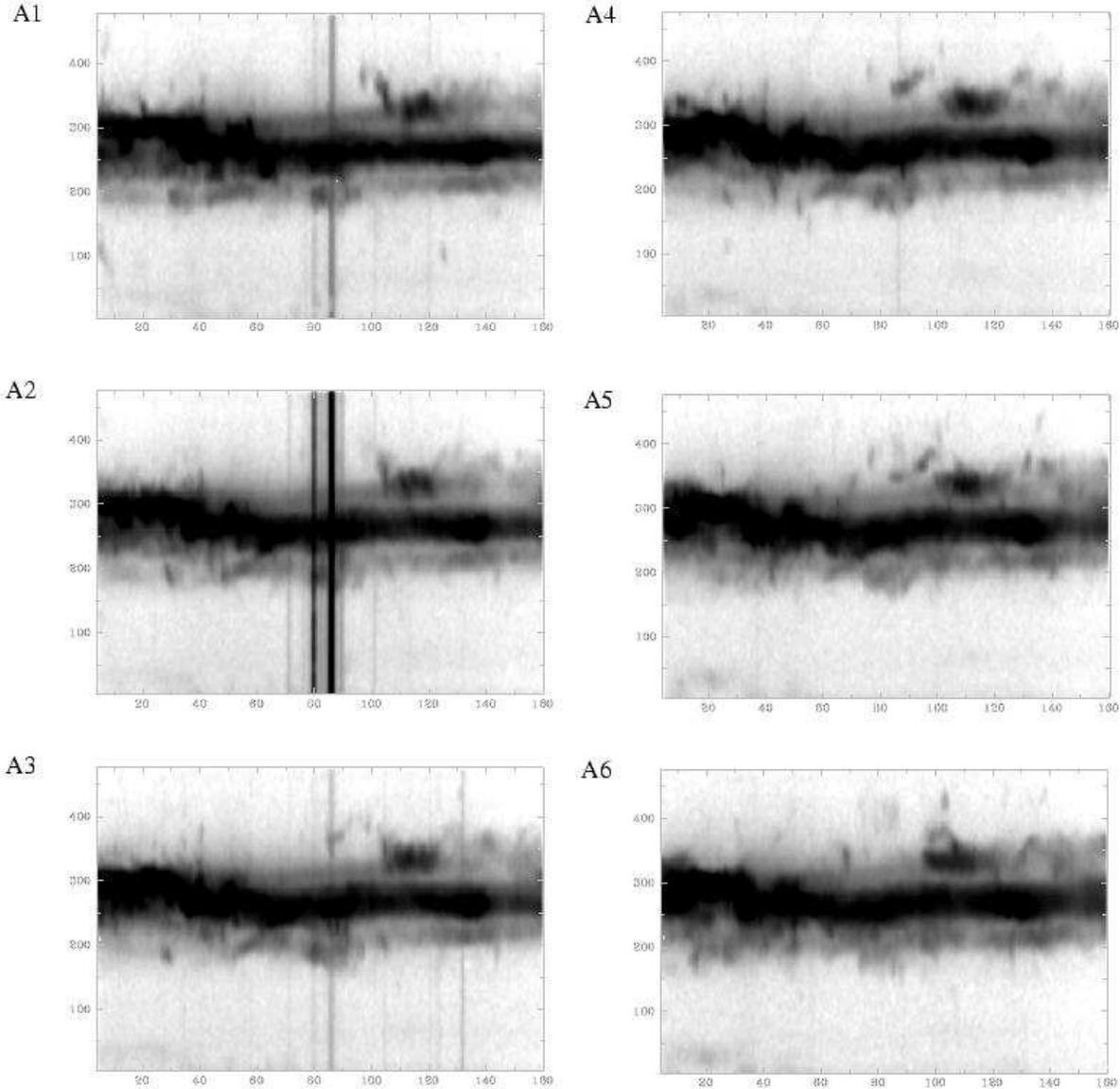,width=450pt,bbllx=20pt,bblly=20pt,bburx=575pt,bbury=563pt}
\caption{Position-velocity arrays of line profiles from slit block A. The
vertical scale is heliocentric radial velocity, $V_{\rm
HEL}~(\rm~km~s^{-1})$ and the horizontal scale is in arcseconds. At the
distance of the LMC, $1 \arcsec=0.27~{\rm pc}$.}
\label{blocka}
\end{figure*}

\begin{figure*}
\psfig{file=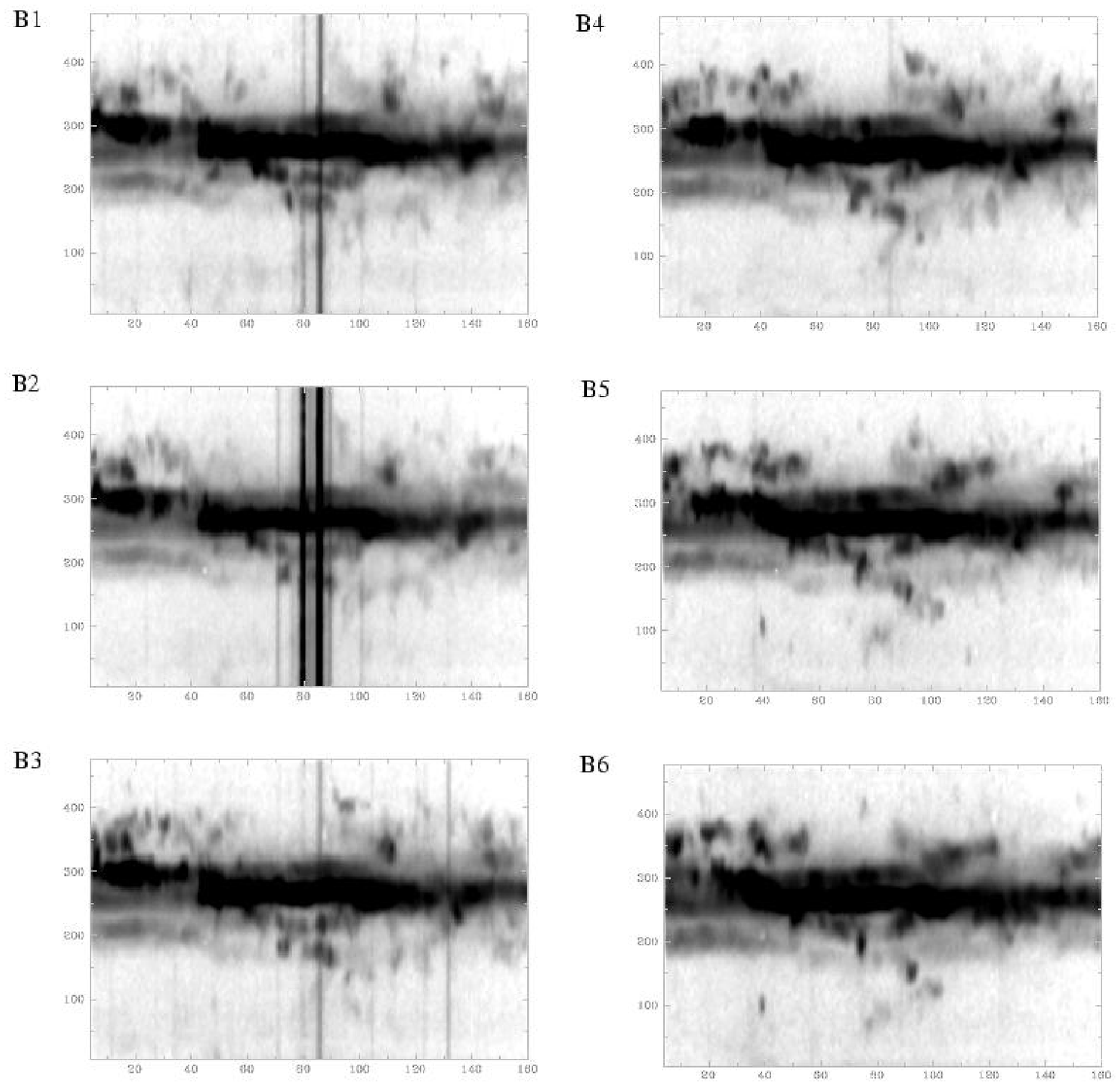,width=450pt,bbllx=20pt,bblly=20pt,bburx=575pt,bbury=562pt}
\caption{Position-velocity arrays of line profiles from slit block B.The
vertical scale is $V_{\rm HEL}~(\rm~km~s^{-1})$ and the horizontal
scale is in arcseconds. At the distance of the LMC, $1
\arcsec=0.27~{\rm pc}$.}
\label{blockb}
\end{figure*}

Greyscale representations of the position-velocity (pv) arrays of line
profiles obtained from all the slits are displayed in
Figs~\ref{blocka}-~\ref{blockc}. These data are also presented and
discussed in Al-Mostafa (1999)\nocite{almostafa99}. The scale is
logarithmic in these three figures to enable all features to be
discerned. In Figs~\ref{deep1} and ~\ref{deep2}, deep representations
of the data from slits C4 and C6 are displayed in order to highlight
the faintest, high velocity material.

These data are amongst the most sensitive obtained for the halo of 30
Doradus and reveal a wealth of faint, complex high speed
features. \citet{chu&kennicutt94} carried out echelle spectroscopy
across the halo of 30 Doradus but many of the faint, highest speed
phenomena revealed here were not detected. In the following
discussion, general trends amongst the wealth of complex kinematical
features are highlighted. Discussion of individual features may be
found in Al-Mostafa (1999). Adopting a distance to the LMC of 55 Kpc
gives that $1''\approx 0.27~{\rm pc}$.

\subsection{Bright systemic features}
The pv arrays are dominated by very bright continuous velocity
features. These are due to the ionized overlapping H~{\sc i} sheets
close to the systemic velocity being disturbed by the slowly expanding
($50~{\rm km~s^{-1}}$) giant shells around 30 Doradus
(\citealt{meaburn80,chu&kennicutt94}). These systemic kinematical
structures have been investigated in detail in earlier work on the
halo of 30 Doradus.

\subsection{Discrete high speed knots}
At the smallest scale (a few arcseconds or approximately one parsec),
the pv arrays contain numerous localised high speed ($\pm 200~{\rm
km~s^{-1}}$ with respect to the $V_{\rm sys}$) velocity knots which do
not appear to vary widely in spatial scale. They are visible in all
the pv arrays and represent the finest-scale high speed substructure
detected here.

\subsection{Velocity loops and arcs}
\begin{figure*}
\psfig{file=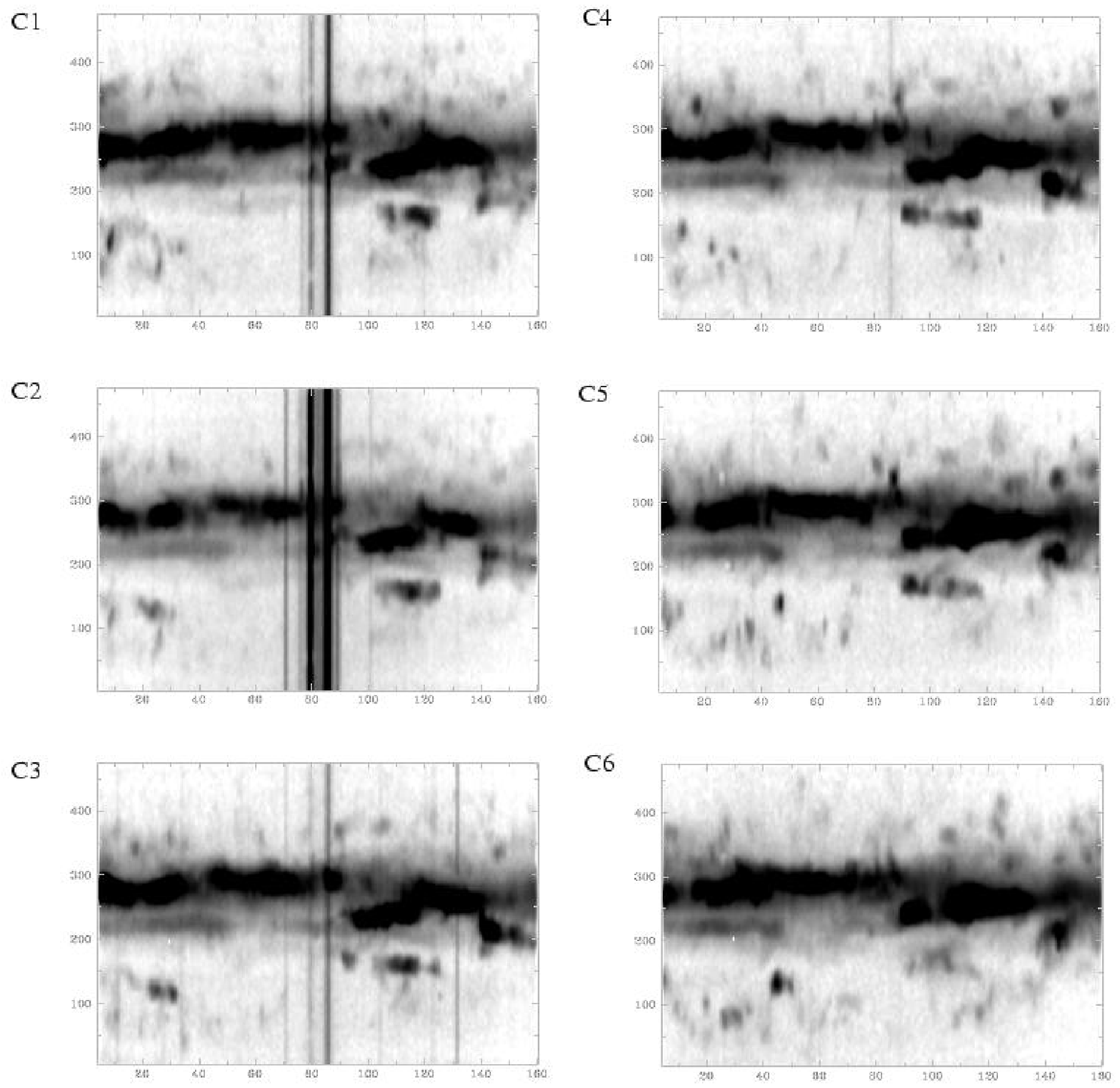,width=450pt,bbllx=20pt,bblly=20pt,bburx=575pt,bbury=561pt}
\caption{Position-velocity arrays of line profiles from slit block C. The
vertical scale is $V_{\rm HEL}~(\rm~km~s^{-1})$ and the horizontal
scale is in arcseconds. At the distance of the LMC, $1
\arcsec=0.27~{\rm pc}$.}
\label{blockc}
\end{figure*}

At a slightly larger scale (around ten arcseconds or a few parsecs),
individual loops and arcs are discerned. They are not continuous in
the pv arrays but are coherent velocity features made up of the
discrete velocity knots. The clearest example is that seen at slit
position C6 (Fig.\@~\ref{blockc} and the deep representation,
Fig.\@~\ref{deep2}).

\subsection{Large scale coherent velocity features}
At the largest scales (tens of arcseconds or approximately ten
parsecs), high speed knots are found to trace out coherent velocity
features that slowly vary between being red and blue shifted. The
clearest example is perhaps that in slit position B3 where at offsets
of approximately 0 to 50\arcsec\ the feature is redshifted and between
around 50 and 100 \arcsec\ it is blueshifted with respect to the
bright continuous feature.

\section{Discussion}

\subsection{Giant shells in 30 Doradus}
The current explanation for the origin of the 30 Doradus nebula is that as
the massive stars within the nebula evolve, their winds (especially
during the Wolf-Rayet phase) and subsequent supernova explosions
generate swept-up shells of ionized gas
\citep{meaburn88,meaburn91}. The shells are observed to have a
hierarchy of sizes and velocities as one moves further into the halo of
30 Doradus. In the dense centre, the shell sizes and velocities are
$\sim 1~{\rm pc}$ and $\sim 10-50~{\rm km~s^{-1}}$ respectively, while
in the halo the sizes reach $\sim 100~{\rm pc}$ with velocities of up
to $100~{\rm km~s^{-1}}$ (see Fig.\@~1). The shells are prone to
instabilities and can break up and fragment, venting the interior
pressure. For example a dense shell that is accelerating (due to
either a rapid drop in the external density or to a new supernova
explosion within the shell) may break up via the Rayleigh-Taylor
instability. Alternatively, dynamical overstabilites can also lead to
a shell breaking up into fragments \citep{maclow&norman93}. In both
cases, the fragments produced will have sizes of the order of the
shell thickness. In general, the halo of 30 Doradus, into which the
shells are expanding, is inhomogeneous. This inhomogeneity, and the
disruptive effects of nearby supernovae and winds from stars not
within the original shell, will mean that a shell will not remain
coherent for long.

It is important to note that the LMC is thought to be flattened and
viewed close to face-on. The scale height of the H~{\sc i} in the LMC
disk was calculated by \citet{kim.et.al99} to be $\sim 180~{\rm pc}$
so that it is likely that the structure 30 Doradus may also be
somewhat flattened. The scale-height imposes a limit on the sizes of
the shells that can be formed, irrespective of how intense and coeval
in time the massive star activity is. As the shells grow, they become
elongated in the direction of the density scale height
\citep{koo&mckee92}, leading to a break up of the shell in this
direction and a `blow-out' of the hot interior gas into the galactic
halo. The remaining structure is known as a galactic chimney
\citep{norman&ikeuchi89} and they have been observed in the Galaxy
\citep{normandeau.et.al96} and in the starburst galaxy M82
\citep{wills.et.al99}.

The giant shells of 30 Doradus are likely to be the maximum sized
spherical momentum conserving shell structures, since the scale height
of the LMC is comparable to their diameters. The supergiant shells far
exceed this scale height and may be collections of fossil chimneys
viewed face-on and also the result of propagating star formation (see
e.\@g.\@~\citealt{mccray&kafatos87}) that is constrained to proceed in
the plane of the galaxy, resulting in a ring shape supergiant
shell. The loss of driving pressure means they are expanding in a
momentum conserving phase and surround a low density cavity. Such
cavities are clearly seen in the H~{\sc i} data of
\citet{kim.et.al99} and \citet{staveley-smith.et.al02}. The hot gas
that has escaped from the interior of the giant shells will enter the
LMC halo. There is strong evidence for such a halo in the
LMC. \citep{wakker.et.al98} have used GHRS/HST observations to detect
C~{\sc iv} absorption towards LMC stars that do not to reside within a
shell. This means the hot gas implied by these observations is not
local to the star and is likely to reside in the halo (see also
\citealt{savage.et.al97}).


\begin{figure*}
\psfig{file=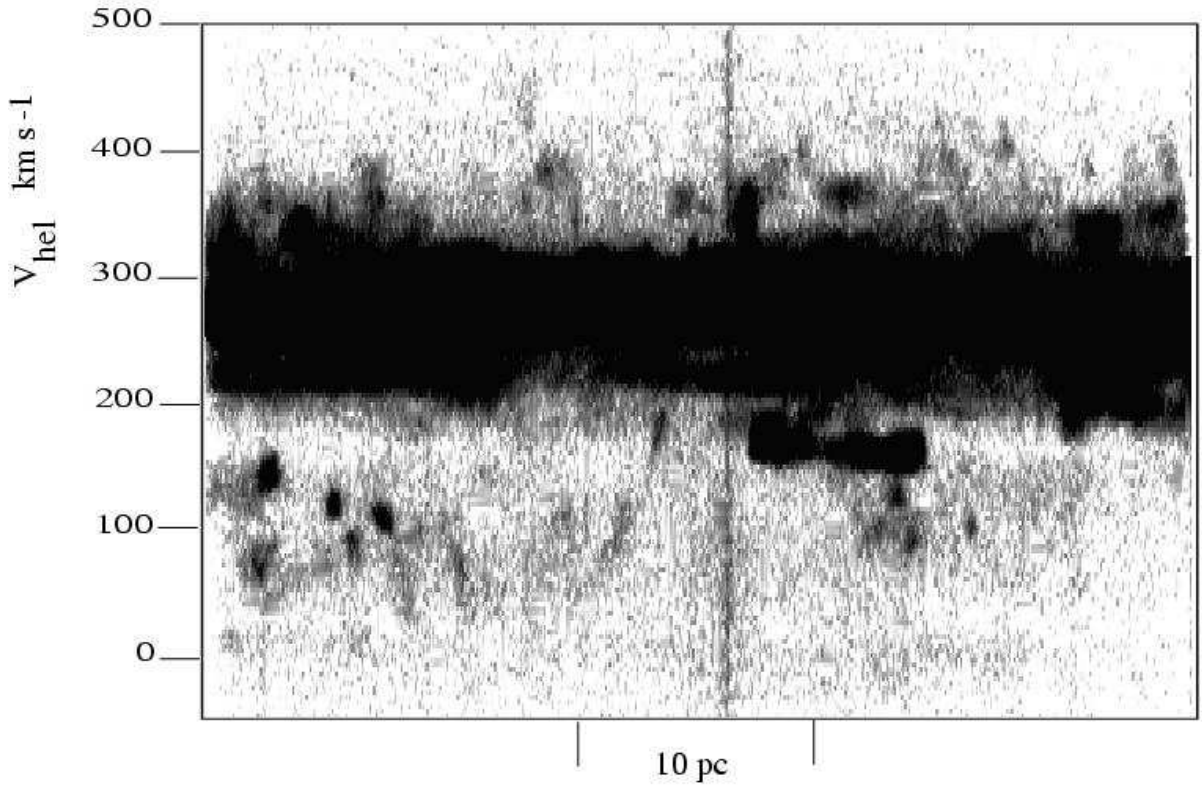,width=450pt,bbllx=20pt,bblly=20pt,bburx=575pt,bbury=389pt}
\caption{Deep presentation of position-velocity arrays of line profiles 
from slit block C4 to highlight fainter features.}
\label{deep1}
\end{figure*}
\begin{figure*}
\psfig{file=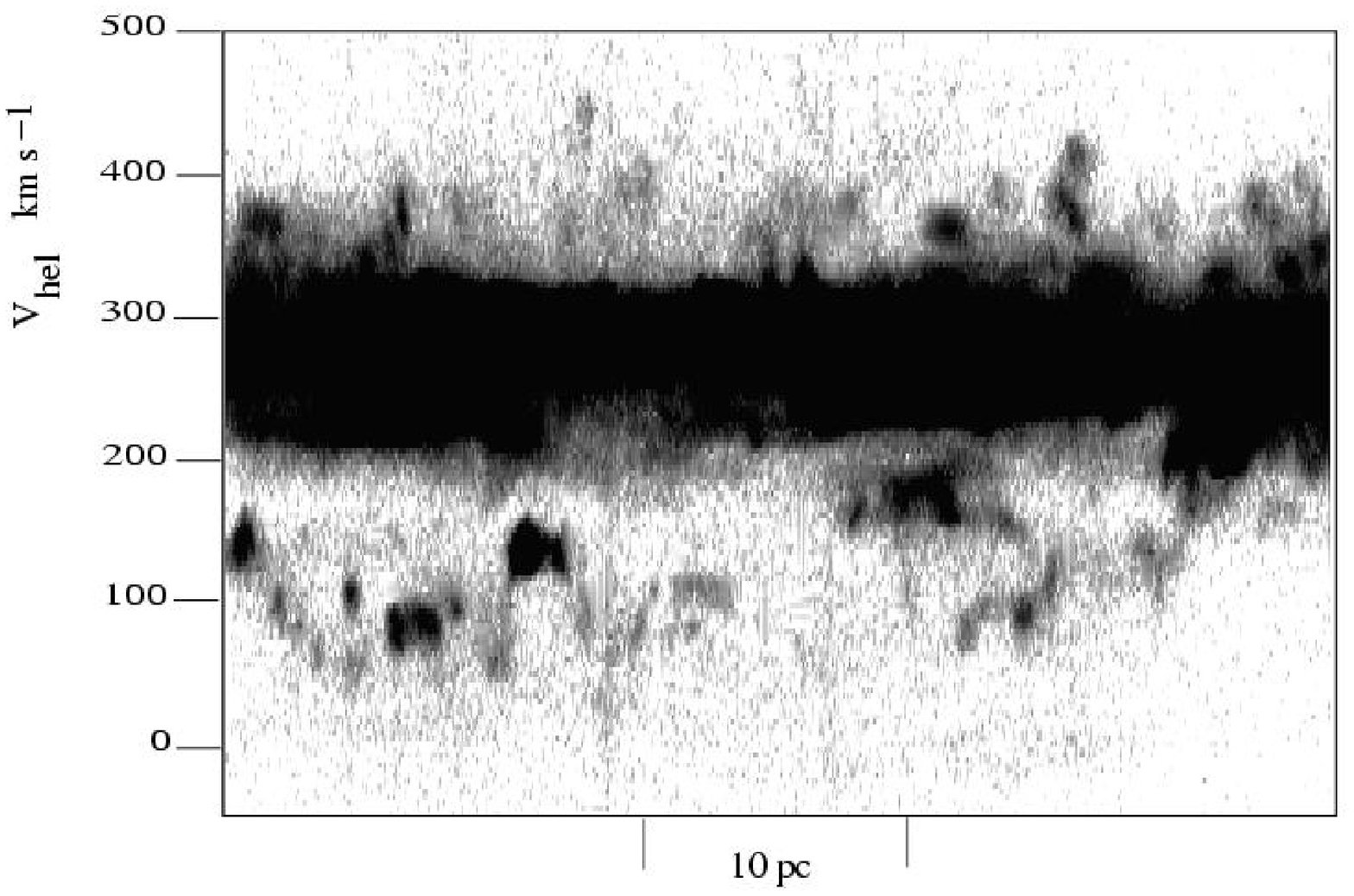,width=450pt,bbllx=20pt,bblly=20pt,bburx=575pt,bbury=387pt}
\caption{Deep presentation of position-velocity arrays of line profiles 
from slit position C6 to highlight fainter features.}
\label{deep2}
\end{figure*}

\subsection{Origin of the velocity features}
It would seem unlikely that the high speed knots represent random gas
clouds within 30 Dor since that would require an explanation for both
their hypersonic velocities and the systematic way they are
distributed about the pv arrays. In terms of the scenario discussed
above (section 3.1), a straight-forward interpretation of the
kinematical features is as follows. The largest scale features
represent old giant shells that have broken up via Rayleigh-Taylor
(RT) instabilities. An instability is generated as the shell is
accelerated by its interior pressure through a decreasing ambient
density. Those portions of the shell that are expanding in the plane
of the LMC are less prone to disruption. The fragments that used to be
part of the shell have continued to coast at the pre-break up velocity
and together these remain as a coherent velocity feature. For the RT
instability, the characteristic knot size at the break up of the shell
will be of the order of the thickness of the shell. The sizes implied
by this picture seem reasonable - the old shell will have a dimension
of up to a hundred parsecs towards the outer regions of the nebulosity
while the shell wall will have had a thickness of a few parsecs. These
estimates are in accord with measurements of intact shells observed
within the halo of 30 Doradus and elsewhere in the LMC
(e.\@g.\@~\citealt{oey96}).

The smaller chains of high-speed knots could be due to more localised
disruptions of the giant shells due to, for example, a
neighbouring supernova explosion. This latter scenario was proposed by
\citet{redman.et.al99b} to explain the unique Honeycomb nebula, which
lies in the halo of 30 Doradus. They argued that its cellular structure is
due to a shell that has begun to fragment by a RT instability being
impacted by a blast wave from a nearby SN explosion. There have been
approximately $\sim$ 40 supernova explosions within the halo of 30 Doradus
in the last $10^4~{\rm yr}$ alone \citep{meaburn91} (in comparison,
in the starburst galaxy M82 there have been $\simeq 50$ SNe in the
last $\simeq 200~{\rm yr}$; \citealt{muxlow.et.al94}).

In the H~{\sc i} pv array data of \citet{staveley-smith.et.al02}, the
LMC and the Galaxy are well separated in velocity. In their data, the
Galaxy does not exhibit kinematical features with a $V_{\rm HEL}\ga
100~{\rm~km~s^{-1}}$ while the LMC does not exhibit kinematical
features with a $V_{\rm HEL}\la 100~{\rm~km~s^{-1}}$. In our data
faint velocity features are seen from the $V_{\rm HEL}$ of 30 Doradus
down to a $V_{\rm HEL}\la 100~{\rm km~s^{-1}}$. However, it is
unlikely that these velocity features are associated with the Galaxy
rather than the LMC for several reasons. Firstly, such features are
not seen at slit positions offset from 30 Doradus; secondly, many of
the features can be traced back to the 30 Doradus systemic velocity;
thirdly, there is no known Galactic \hii\ region or ionizing source
along the line of sight that could excite the \oiii\ emitting gas and
there is also no extensive background
\oiii\ emission in the galaxy (compare Fig.\@~\ref{deep1} and~\ref{deep2} 
here with figure 9 of \citealt{staveley-smith.et.al02}).

\subsection{Superwind from 30 Doradus}
Starburst activity can give rise to a `superwind' due to the intense
radiation fields, winds and supernova explosions caused by the star
formation. The extensive high velocity features revealed here may be
marking the very base of a superwind localised around the 30 Doradus
complex. The escape velocity of gas from the LMC in the neighbourhood
of 30 Doradus is around $150~{\rm km~s^{-1}}$ so that the high speed
ionized gas from disrupted shells and giant shells is escaping the
gravitational pull of 30 Doradus and is being ejected perpendicularly
to the plane of the LMC, along the line of sight. The ionization
boundary due to the R130 cluster (perpendicular to the plane of the
LMC) will depend on the distribution of the gas but an upper limit of
a few hundred parsecs can be estimated by calculating the Str\"omgren
radius due to an ionizing flux of $\sim 10^{51}~{\rm s^{-1}}$ from the
$\sim 100$ O stars and a mean gas density of $\sim 1~{\rm
cm^{-3}}$. Assuming an ejection speed of $200~{\rm km~s^{-1}}$, it
will take of the order $2\times 10^6~{\rm yr}$ to reach a distance of
$\sim 500~{\rm pc}$ from the point of origin and thus escape the 30
Doradus region. The gas will rapidly recombine once it has passed the
ioniziation boundary and will then not be visible on the \oiii\ pv
arrays. The dynamical timescale of the remaining giant shell walls is
much longer since their progress in the direction of the plane of the
LMC is slower than the material ejected perpendicular to the
plane. High velocity H~{\sc i} clouds will be formed by the escaping
material as it recombines and these clouds may be detectable in high
resolution and high sensitivity H~{\sc i} kinematical studies. Of
course, the kinematics of gas ejected from the LMC rapidly becomes
highly complex due to the interaction with the Galaxy
\citep{wakker&woerden97}.

\section{Conclusions}
In this work the kinematics of sample region of the 30 Doradus nebula
have been investigated using the MES. This intensive study has
revealed high speed velocity features throughout this region.
Although the kinematics are complex, general patterns are discerned at
three different spatial scales. Small coherent velocity features are
present throughout the region. These knots are often found to form
loops and chains in the pv arrays and at the largest scales, can form
velocity features which vary slowly between red and blue-shifted
emission. It is suggested that all of these features are explicable in
terms of the current understanding of the 30 Doradus nebula. Shells
and giant shells formed by the winds and supernovae of massive stars
form and are then disrupted in the energetic turbulent environment of
the halo of 30 Doradus. The fragments of the shells retain the
velocity pattern of the original shell and are observed as the small
high speed knots. If this explanation is correct, then high velocity
knots are likely to be found across much of the face of 30 Doradus
wherever the size of the giant shells have exceeded the scale-height
of the LMC and led to `blow-out'. The whole 30 Doradus nebula is
flattened and viewed face on. The high speed velocity fragments are
likely to form the base of an outflowing superwind that is escaping
the galaxy. This is a microcosm of the processes that are taking place
in starburst galaxies such as M82 in which there are many super star
clusters like 30 Doradus and whose combined output lead to the
spectacular optical filaments that mark the M82 superwind.

\section*{Acknowledgements}
JM and MB would like to thank the staff at the AAT, who provided their
usual excellent service during the observing run. MPR is supported by
PPARC. A King Abdulaziz City for Science and Technology `KACST'
studentship is acknowledged by ZAA. We thank the referee for comments
which improved the paper.

\label{lastpage}
\end{document}